\documentclass[floatfix,aps,twocolumn,showpacs,nofootinbib,superscriptaddress]{revtex4}

\usepackage{graphicx}
\usepackage{subfigure}
\usepackage{epsfig}
\usepackage{amsmath}
\usepackage{amsfonts}
\usepackage{amssymb}
\usepackage{braket}
\usepackage{hyperref}
\usepackage{dsfont}
\usepackage{qcircuit}
\usepackage{color}
\definecolor{darkgreen}{RGB}{40,130,40}
\definecolor{darkblue}{RGB}{50,50,200}

\begin{document}
\author{Nahuel Freitas}
\email{nahuel.freitas@physik.uni-saarland.de}
\author{Giovanna Morigi}
\affiliation{Theoretische Physik, Universit\"{a}t des Saarlandes,
D-66123 Saarbr\"{u}cken, Germany}
\author{Vedran Dunjko}
\affiliation{Max Planck Institut f\"{u}r Quantenoptik, Hans-Kopfermann-Str. 1,
85748 Garching, Germany}

\title{Neural Network Operations and Susuki-Trotter evolution of Neural Network States}

\date{\today}

\begin{abstract}
It was recently proposed to leverage the representational power of artificial
neural networks, in particular Restricted Boltzmann Machines, in order to model
complex quantum states of many-body systems [Science, 355(6325), 2017].
States represented in this way, called Neural Network States (NNSs), were shown
to display interesting properties like the ability to efficiently capture
long-range quantum correlations.
However, identifying an optimal neural network
representation of a given state might be challenging, and so far this problem has been
addressed with stochastic optimization techniques. In this work we explore
a different direction.
We study how the action of elementary quantum
operations modifies NNSs. We parametrize a family of many body quantum operations
that can be directly applied to states represented by Unrestricted Boltzmann Machines,
by just adding hidden nodes and updating the network parameters. We show that this
parametrization contains a set of universal quantum gates, from which it follows
that the state prepared by any quantum circuit can be expressed as a Neural Network State
with a number of hidden nodes that grows linearly with the number of elementary
operations in the circuit. This is a powerful representation theorem (which
was recently obtained with different methods) but that is not directly useful,
since there is no general and efficient way to extract information from this
unrestricted description of quantum states. To circumvent this problem,
we propose a step-wise procedure based on the projection of
\emph{Unrestricted} quantum states to \emph{Restricted} quantum states.
In turn, two approximate methods to perform this projection are discussed.
In this way, we show that it is in principle possible to approximately optimize or
evolve Neural Network States without relying on stochastic methods
such as Variational Monte Carlo, which are computationally expensive.
\end{abstract}

\maketitle

\section{Introduction}

As is well known, the description of general quantum states of composite
systems requires an amount of information that grows exponentially with the
number of subsystems. This simple fact is one of the reasons why general
quantum systems are hard to simulate with ordinary computers.
A possible workaround for this problem is to abandon the desire of
describing arbitrary quantum states, and only concentrate on a manifold of
physically meaningful states\cite{poulin2011}.
A prominent example along this line is given by the Matrix Product States (MPSs)\cite{orus2014}
\footnote{Here, we assume a restricted bond dimension of the MPS, such that the total number of parameters which need to be specified is polynomial in the number of subsystems. If the bond dimension is not restricted, any
quantum state can be cast as an MPS.}.
Here, the physically meaningful states that are addressed are the low energy
states of gapped Hamiltonians with local interactions. In one dimension, it is
known that those states satisfy an entanglement area law, and MPSs are a
sufficiently general class of states compatible with such law
\cite{xiang2001,legeza2004,vidal2003,srednicki1993,plenio2005}.
This is also a limitation for MPSs,
since they are then not sufficient to efficiently capture the rich physics close to quantum
critical points, where the gap typically closes and the quantum
correlations no longer obey an area law\cite{wolf2006,kallin2011}. This is also
the case for systems with long range interactions\cite{dogra2016,landig2016,niederle2016,deng2017}.

\begin{figure}
  \centering
  \includegraphics{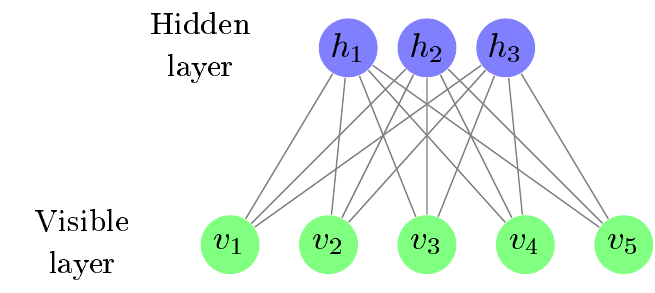}
  \caption{Restricted Boltzmann Machine with $N=5$ visible and $M=3$ hidden nodes.
  When modelling a quantum state, each of the visible nodes $v_1,\cdots,v_N$
  represents, for instance, a spin $s_j$ on a lattice. The number of hidden neurons
  determines the power that the network has to represent distributions over
  the visible nodes.}
  \label{fig:rbm}
\end{figure}

Recently, a new family of states was proposed by Carleo and
Troyer\cite{carleo2017} to deal with long range quantum correlations in
many-body systems: the so-called Neural Network States (NNSs) or Neural Quantum
States (NQSs). The main idea
behind this proposal is to treat the wave function as a functional that maps
configurations of lattice spin systems (states of a given computational basis)
to complex numbers (probability amplitudes). As the name suggests, a neural
network architecture is used to model this mapping. In particular, the
neural networks employed in \cite{carleo2017} are Restricted Boltzmann Machines (RBMs).
An RBM is defined in terms of a network of hidden and visible nodes, with
weighted connections between these two groups, which thus form a bipartite
graph (see Figure \ref{fig:rbm}). An Ising-like energy functional is assigned
to the network, and the distribution realized by it is specified by the
Boltzmann factor corresponding to that energy, often conditioned on some
configuration of either the visible or both hidden and visible parts of the
network. The `programming' or `training' of the network consists in adjusting
the weights such that the network matches, or well approximates, a target distribution
over the visible nodes, which is often specified only via a set of samples drawn from it.

In order to model a quantum many body state each visible node is associated
with a subsystem, for example a spin on a lattice, and the expressive power\footnote{By expressive power here we mean the size and complexity of the set of distributions which can be realized over the visible nodes, by marginalizing over the hidden nodes.},
of the machine increases with the number of hidden nodes considered.

The family of states thus obtained with efficient NNSs, that is, those utilizing only a polynomial number of hidden nodes was later shown to be able to efficiently describe
volume-law entanglement\cite{deng2017}\footnote{It is well-known that RBMs (and NNSs) can represent any distribution and/or state, provided the number of hidden nodes is not limited. This is similar to how unlimited bond dimensions render MPSs fully expressible.}. The relationship between NNSs and MPSs
was explored in \cite{chen2017}, showing that, in general, in order to exactly represent
a given NNS as a MPS, an exponentially large bond dimension is required for the latter.
Thus, it was shown that MPSs cannot efficiently describe general
NNSs based on RBMs (RBM-NNSs). In subsequent works, it was shown that RBM-NNSs are also related to other previously known families such as String-Bond States\cite{glasser2018}, or arbitrary graph states\cite{gao2017}.

So far, NNSs were mainly employed as a variational Ansatz, either to minimize the
energy of a model Hamiltonian, to evolve a state over time, or for quantum state
tomography\cite{deng2017,glasser2018,carleo2017,torlai2017}. In these cases
the neural network representing the state was optimized by usual Variational Monte
Carlo (VMC) techniques. This is computationally expensive, since at each iteration in
the optimization it is necessary to stochastically estimate the gradient
of an objective function with respect to the network parameters\cite{carleo2017}.
In this work we explore a different direction. Our main aim is to find a
method to evolve a given NNS, on the level of the representation and
in a controlled way, without requiring stochastic sampling and estimation.
We begin by investigating how to update the parameters of a given NNS
to take into account the action of simple physical processes.
Thus, we pose the following question: is there any family of
(non-trivial) elementary quantum operations (\emph{i.e.} unitary gates) that can be applied to a NNS state,
in such a way that the resulting state is also efficiently representable as a NNS, and such that the update can be performed on the level of the NN representation itself?
How can this family be parametrized?
An answer to this question would shed further light on the properties and
limitations of NNSs, and could guide the development of efficient numerical
algorithms to evolve and optimize NNSs, without the need to rely exclusively
on stochastic methods.
In other words, what we are asking for is a collection of efficient rewriting
rules, which can approximate the evolution of quantum states, on the level of
the graphs representing them.

As it turns out, even very simple one-body unitary operations take
a general RBM-NNS to a new state that is naturally represented
by \emph{Unrestricted} Boltzmann Machines (UBMs), for which connections
among the hidden nodes are allowed. We point out that this does not mean that applying a single-body
unitary to a RBM-NNS necessarily renders this state into a new one without an efficient
RBM representation. However, an extremely simple update rule can be identified
if the restriction on the connections in the hidden layer is relaxed.
We will refer to states described by UBMs as UBM-NNSs.
For this more general representation, there exists
a family of non-trivial operations whose action can be easily represented by
simple update rules.
Specifically, we identify a family of $K$-body operations that
can be applied to any UBM-NNS by only adding $K$ hidden nodes to the UBM,
followed by a simple update of the network parameters.
This family contains universal sets of gates, so in this way, we also show that
the quantum state prepared by any quantum circuit can be expressed as
a UBM-NNS with a number of hidden units that increases linearly with the number
of elementary operations in the circuit, provided that the initial state is
also a UBM-NNS.
These results are compatible to those obtained in \cite{gao2017},
where the representational power of \emph{Deep} Boltzmann Machines was explored
and compared to the shallow or restricted case\footnote{As explained in
\cite{gao2017}, Deep and Unrestricted Boltzmann machines are essentially equivalent.}.
However, our methods are different and provide new insights.
Although the mentioned results are interesting and show the power of
Boltzmann Machines to represent quantum states, they are not directly useful.
The reason is that, in contrast to RBM-NNSs, there is no accurate and
efficient way to extract information out of a UBM-based description
of a quantum state. To explain this we compare the problem of sampling a UBM-NNS
to standard Quantum Monte Carlo techniques based on path integrals
where a classical model, \emph{`dual'} to a quantum model of
interest, is sampled\cite{ceperley1986, sachdev2007, troyer2005}.

In fact, the main advantage of RBMs is that they can be sampled efficiently
(since the quotients between probability amplitudes can be readily computed).
Of course, this comes at the expense of some representational power\cite{gao2017}.
Nevertheless, as mentioned before, RBM-NNSs can still represent many complex
and highly entangled quantum states\cite{deng2017,chen2017,gao2017,glasser2018}.
Thus, building on the study of quantum operations, we propose a method to
evolve an initial RBM-NNS
in such a way that the final state is also an easy to sample RBM-NNS.
The central idea is that, whenever a
quantum operation transform the input state to a UBM-NNS, this output state
is \emph{projected} back to the family of RBM-NNS.
Two projection
procedures are presented and discussed.
Finally, this ideas are tested showing
that it is possible to optimize RBM-NNSs in order to approximate the ground
state of the transverse field Ising model in one dimension without employing
stochastic methods like VMC. To the best of our knowledge, this is the first
example of a method in which RBM-NNS are optimized in a deterministic manner,
providing an alternative to stochastic methods.

This article is organized as follows: in section \ref{sec:one-body} we
review the definition of RBM-NNS and show how the action of simple one-body
unitaries takes them to UBM-NNS. In section \ref{sec:two-body} we define
general UBM-NNS and show how the network must me modified to take into account
the action of K-body operations. In section
\ref{sec:sampling} we compare the sampling of UBM-NNS to usual QMC methods.
In section \ref{sec:projection} we propose a method to continuously project
the evolved state back to the family of RBM-NNS. Finally, in Section
\ref{sec:optimization} we apply this ideas and show how to approximate the
ground state of an Ising chain.

\section{RBM-NNS\lowercase{s} and one-body operations}
\label{sec:one-body}

Boltzmann machines are described by a set of nodes (neurons), representing stochastic
variables, with bidirectional weighted connections between them forming a neural network.
These nodes are usually split in two groups: visible nodes and hidden nodes.
The restriction in Restricted Boltzmann Machines (RBMs) is that no connections are
allowed between nodes of the same group, as depicted in Fig. \ref{fig:rbm}.

In classical RBMs an energy function is assigned to the network, which is
typically a quadratic function of the node values:
\begin{equation*}
  E_{\text{RBM}} = -a^t v - b^t h - h^t W v,
\end{equation*}
where $v=(v_1,\cdots,v_N)^t$ and $h=(h_1,\cdots,h_M)^t$ are column vectors with
the values of the $N$ visible nodes and the $M$ hidden nodes, respectively. The
vectors $a=(a_1,\cdots,a_N)^t$ and $b=(b_1,\cdots,b_M)^t$, along with the
$M\times N$ matrix $W$, are the parameters of the network. The constants $a_k$ and
$b_k$ are known as offsets, and the components of the matrix $W$ indicate the weights
of the connections. In classical
applications the parameters $a$, $b$, and $W$ are real, and it is assumed that
the probability distribution for the stochastic variables in $h$ and $v$,
$P(v,h|a,b,W)$, is of the Boltzmann form:
\begin{equation*}
  P(v,h|a,b,W) = \frac{1}{Z} e^{-E_\text{RBM}}
\end{equation*}
where $Z=\sum_{v,h} e^{-E_\text{RBM}}$ is the partition function.
Now, this network can be used to learn a target probability distribution over
the visible nodes. Given a training set of configurations for the visible nodes,
with a distribution $P_T(v)$, training the network means to adjust the network
parameters $a$, $b$, and $W$ in order to minimize some measure of
distance\footnote{For example, the Kullback-Leibler divergence\cite{leroux2008}.}
between the marginal distribution over the visible nodes
\begin{equation}
P(v) = \sum_{h_1,\cdots,h_M} P(v,h|a,b,M)
\label{eq:RBM_sum_hidden}
\end{equation}
and $P_T(v)$.
This is an optimization problem that can be attacked with different iterative
methods, such as so-called contrastive divergence\cite{bengio2009}.
The fact that hidden nodes are not connected between them in RBMs (i.e, the
energy is only linear in $h$), allows to explicitly perform
the sum to find $P(v)$. If it is assumed for simplicity that the hidden variables
$h_k$ are binomial, taking the values $\{-1,1\}$, then from
Eq. (\ref{eq:RBM_sum_hidden}) we obtain:
\begin{equation*}
  P(v) \propto e^{a^t v} \prod_{m=1}^M \cosh(b_m + (W v)_m),
\end{equation*}
apart from a normalization constant.

As proposed in \cite{carleo2017} it is possible to extend this approach to
learn, or model, a quantum state or wavefunction instead of a probability
distribution. For this, let us consider a many body system with
$N$ subsystems, each with $R$ levels $\{\ket{s}\}_{s=1,\cdots,R}$.
A computational basis for the whole system can be given by the product
states $\ket{s_1,s_2,\cdots,s_N} = \otimes_{k=1}^{N} \ket{s_k}$. Any many-body
state $\ket{\Psi}$ can then be considered as the mapping
$\Psi(s_1,\cdots,s_N) \equiv \braket{s_1,\cdots,s_N | \Psi}$
of each of these product states to a complex probability amplitude.
Thus, by associating each variable $s_k$ with a
visible node of a RBM, we can model this mapping as:
\begin{equation}
\begin{split}
  \Psi(s) \propto \sum_{h_1,\cdots,h_M} e^{a^t s + b^t h + h^t W s}
\label{eq:rbm_nns}
\end{split}
\end{equation}
where $s=(s_1,\cdots,s_N)^t$, and the parameters $a$, $b$, and $W$ are allowed
to be complex. As in the previous equation, in the rest of this work we will
describe quantum states up to an unspecified normalization constant. It is not
required to known this constant since in order to estimate the expectation
value of physical quantities for a given state we employ stochastic methods
which only need to evaluate the ratios $\Psi(s)/\Psi(s')$ for different
configurations $s$ and $s'$, as explained in detail in \cite{carleo2017}.

In the following we will focus in
the case in which each subsystem is a two-level system (i.e, a spin-1/2 or a qubit),
so that the visible nodes are also binomial variables and each $s_k$ can only
take the values $\{-1,1\}$ (i.e, $R=2$). We will refer to quantum states written
as in Eq. (\ref{eq:rbm_nns}) as a RBM-NNS.

\subsection{Action of one-body operations}

\begin{figure}
  \centering
  \includegraphics{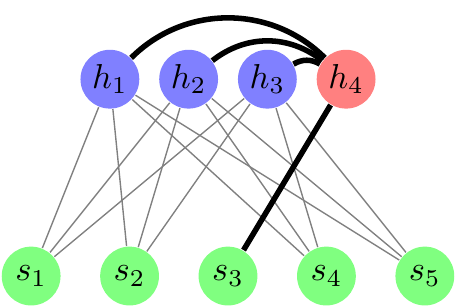}
  \caption{Unrestricted Boltzmann machine with 5 visible units and 4 hidden units,
  resulting from applying a one-body operation to the spin represented by the visible
  node 3 of Fig. \ref{fig:rbm}. The new hidden node is colored in red and the new
  connections are marked by thick lines.}
  \label{fig:one_body_ubm}
\end{figure}

Let us consider a linear operation $U^{(j)}$ acting on the Hilbert space of
subsystem $j$ (a single spin-1/2). How does this operation act on a given RBM-NNS?
First, we note that if $U^{(j)}_{s,s'} = \bra{s}U^{(j)}\ket{s'}$ are the matrix elements
of $U^{(j)}$, and $\Psi(s)$ is an arbitrary wavefunction, then the wavefunction
corresponding to the state $\ket{\Psi'} = U^{(j)} \ket{\Psi}$ is:
\begin{equation*}
  \Psi'(s)=\sum_{s'_j} \; U^{(j)}_{s_j,s'_j} \; \Psi(s_1,\cdots,s_{j-1},s'_j,s_{j+1},\cdots,s_N).
\end{equation*}
We will assume for the moment that the matrix elements of the operation $U^{(j)}$ can be
expressed as:
\begin{equation}
  U^{(j)}_{s,s'} = A\: e^{\alpha s + \beta s' + \omega s s'},
  \label{eq:simple_one_body}
\end{equation}
with complex parameters $\alpha$, $\beta$ and $\omega$. If the operation
$U^{(j)}$ is required to be unitary, then the value of the constant $A$ should be
such that $\det(U^{(j)}) = 1$. As we explain later, up to a global phase,
any spin-$1/2$ unitary can be described in this way. However, this
parametrization also allows for non-unitary operations.
Then, if the initial state $\ket{\Psi}$ is a RBM-NNS with $M$ hidden variables
and parameters $a$, $b$ and $W$, we have:
\begin{equation*}
\begin{split}
  \Psi'(s)&\!= \!\!\!\!\!\!\!\!\! \sum_{h_1,...,h_M, s'_j} \!\!\!\!\!\!\!\!
  \exp\!\!\left(\sum_{n\neq j} a_n s_n + \sum_i b_i h_i +
  \sum_i \sum_{n\neq j }h_i W_{i,n} s_n\right)\\
  &\times A \exp\left(a_j s'_j + \sum_i h_i W_{i,j} s'_j +
  \alpha s_j + \beta s'_j + \omega s_j s'_j \right).
\end{split}
\end{equation*}
This last expression is already written in a form that suggest us to consider
the sum index $s'_j$ as a new hidden node. Indeed,
$\Psi'(s)$ can be expressed as:
\begin{equation}
  \Psi'(s) = A \!\!\!\!\!\sum_{h_1,...,h_{M\!+\!1}}\!\!\!\!\! e^{\tilde{a}^t s
  +\tilde{b}^t\tilde{h}+\tilde{h}^t\tilde{W}s+\tilde{h}^t\tilde{X}\tilde{h}/2},
  \label{eq:one_body_ubm_nns}
\end{equation}
which is similar to Eq. (\ref{eq:rbm_nns}) but with an additional term
$\tilde{h}^t\tilde{X}\tilde{h}/2$ describing interactions between hidden
nodes. In the previous expression, the updates to the original vectors
$a$, $b$ and $h$, denoted $\tilde{a}$, $\tilde{b}$ and $\tilde{h}$, are
\begin{equation}
\begin{split}
  \tilde{a} &= (a_1,\cdots,a_{j-1},\alpha,a_{j+1},\cdots,a_N)^t,\\
  \tilde{b} &= (b_1,\cdots,b_M,\beta + a_j)^t,\\
  \tilde{h} &= (h_1,\cdots,h_M,h_{M\!+\!1})^t,
\end{split}
\end{equation}
and the new matrices $\tilde{W}$ and $\tilde{X}$ are given by
\begin{equation}
\begin{split}
  \tilde{W} &=
  \begin{pmatrix}
  | &  & | & 0 & | &  & |\\
  W_{i,1} & \cdots & W_{i,j-1} & \vdots & W_{i,j+1} & \cdots & W_{i,N}\\
  | &  & | & 0 & | &  & |\\
  0 & \cdots & 0 & \omega & 0 & \cdots & 0
  \end{pmatrix},\\
  \tilde{X} &=
  \begin{pmatrix}
    0       &  \cdots &        &    0    & W_{1,j} \\
    \vdots  &  \ddots &        &  \vdots & W_{2,j} \\
            &         &        &         & \vdots \\
    0       &  \cdots &        &    0    & W_{M,j} \\
    W_{1,j} & W_{2,j} & \cdots & W_{M,j} & 0
  \end{pmatrix}.
\end{split}
\end{equation}
Therefore, the state $\ket{\Psi'} = U^{(j)} \ket{\Psi}$ can be
written as a NNS with one more hidden variable with respect to $\ket{\Psi}$
and, more importantly, in terms of an \emph{Unrestricted} Boltzmann machine
(UBM) where the interaction between hidden variables is described by the
matrix $\tilde{X}$. Figure \ref{fig:one_body_ubm} illustrates
the new network giving rise to $\ket{\Psi'}$,
Eq. (\ref{eq:one_body_ubm_nns}),
as if a one-body operation was applied to the subsystem 3 in
Fig. \ref{fig:rbm}. We see that the new hidden node only connects to the
visible node corresponding to the subsystem where the operation was applied,
and also, in principle, to all the preexistent hidden nodes.

We now turn to analyze the family of linear operations given by
Eq. \ref{eq:simple_one_body}. First, we note that for $U^{(j)}$
to be unitary $\alpha$ and $\beta$ must be imaginary, and
$\text{Im}(w)= \pi(n+1/2)/2$ for any integer $n$ (we take $n=0$ in what follows).
Thus, we can rewrite Eq.
(\ref{eq:simple_one_body}) as:
\begin{equation}
  U^{(j)}\! =\! \frac{e^{i\pi/4}}{\sqrt{2 \cosh(2\omega'\!)}}\!
  \begin{pmatrix}
   e^{i(\alpha'+\beta')}e^{\omega'}&-ie^{i(\alpha'-\beta')} e^{-\omega'}\\
   -ie^{-i(\alpha'-\beta')}e^{-\omega'}&e^{-i(\alpha'+\beta')}e^{\omega'}
  \end{pmatrix}.
  \label{eq:one_body_matrix}
\end{equation}
In this form, the new parameters $\alpha'$, $\beta'$ and $\omega'$ are real numbers.
Up to a global phase, Eq. \ref{eq:one_body_matrix} is equivalent to any
spin-$1/2$ unitary operation. It is particularly interesting to analyze the
case of operations that are diagonal in the computational basis (rotations
around $\hat z$). We see that such operations are recovered in the limit
$\omega' \to +\infty$. However, in that case the new hidden node $h_{M+1}$
can be identified with $s_j$ and eliminated (only the terms in which
$h_{M+1} = s_j$ survive when the sum in Eq. \ref{eq:one_body_ubm_nns}
is performed). Thus, rotations around the $\hat z$ axis of subsystem $j$ can be
implemented without adding new hidden nodes, and just updating the value of $a_j$
according to the rule $a_j \to a_j - i\theta/2$, where $\theta$ is the rotation
angle.
We also note that the representation of infinitesimal operations requires large
values of $\omega'$. As an example, for the infinitesimal rotation
$U^{(j)}=\mathds{1} - (i\theta/2) \sigma_j^x$ we have
$\omega'= (-1/2) \log(\theta/2)$. This will be relevant for the analysis of the
projection method presented in Section \ref{sec:projection}.

\subsection{UBM-NNS\lowercase{s} and K-body operations}
\label{sec:two-body}

The results from the previous section motivate us to define an extended
family of NNSs, in which the wavefunction is represented in terms of an
Unrestricted Boltzmann Machine (UBM). In this case, internal connections
in the groups of hidden and visible nodes are allowed. Then, we consider
wavefunctions that can be written as:
\begin{equation}
\begin{split}
  \Psi(s) &= \sum_{h_1,\cdots,h_M} e^{a^t s + b^t h
  + h^t W s + h^t X h/2 + s^t Y s/2},
\label{eq:ubm_nns}
\end{split}
\end{equation}
where the vectors $a$, $s$, $b$ and $h$, and the matrix $W$ are defined as
before, while the symmetric matrices $X$ and $Y$ contain the weights of the
connections within the hidden and visible layer, respectively. They also have
null diagonals (any non zero diagonal element on $X$ on $Y$ will not have any
effect if all the nodes can only take the values $\pm 1$). We will refer to
states written in this way as UBM-NNS.

We are interested in finding a family of linear operations that can be
efficiently applied to the previous states. We will first consider two-body
operations, acting on subsystems $j$ and $k$, such that their matrix elements
$U^{(j,k)}_{rs,r's'} = \bra{r,s} U^{(j,k)} \ket{r',s'}$ can be expressed as:
\begin{equation}
U^{(j,k)}_{rs,r's'} \!=\! A \exp\left(\!
\alpha^t q
\!+\!
\beta^t q'
\!+\! \frac{1}{2} \!
\begin{pmatrix}q^t & q'^t\end{pmatrix} \!\!
\begin{pmatrix}\begin{smallmatrix}0&\lambda \\ \lambda &0\end{smallmatrix} & \Omega \\
\Omega^t & \begin{smallmatrix} 0 & \gamma \\ \gamma & 0\end{smallmatrix}
\end{pmatrix} \!\!
\begin{pmatrix}q \\ q'\end{pmatrix}\!\!
\right)
\label{eq:simple_two_body}
\end{equation}
where $q=\left(\begin{smallmatrix}r \\ s\end{smallmatrix}\right)$,
$q'=\left(\begin{smallmatrix}r' \\ s'\end{smallmatrix}\right)$,
$\alpha =\left(\begin{smallmatrix}\alpha_1 \\ \alpha_2 \end{smallmatrix}\right)$,
$\beta =\left(\begin{smallmatrix}\beta_1 \\ \beta_2 \end{smallmatrix}\right)$,
$\lambda $ and $\gamma$ are constants, and $\Omega$ is a $2\times 2$ matrix.
The parameters $\alpha$, $\beta$, $\lambda$, $\gamma$ and $\Omega$
can in principle be complex valued. In the next section we explain that, at
variance with the single qubit case, not any unitary over two qubits can be written
in this way.

\begin{figure}
  \centering
  \includegraphics{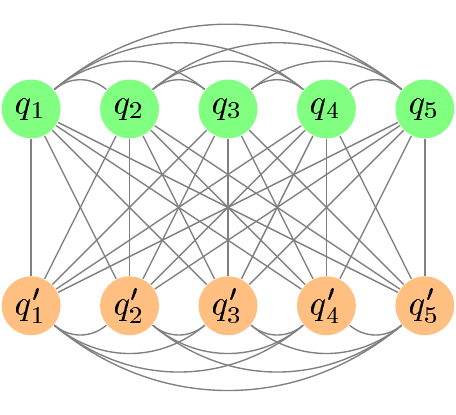}
  \caption{Neural network representing an arbitrary five-body NNO. For given values
  of $q_1,\cdots,q_N$ and $q'_1,\cdots,q'_N$, evaluation of the network gives
  the complex matrix element $\bra{q_1,...q_K} U \ket{q'_1...q'_K}$.}
  \label{fig:nqo}
\end{figure}

As in the previous section,
it can be seen that if the wave function of the state $\ket{\Psi}$ is given
by Eq. (\ref{eq:ubm_nns}), then the wavefunction of $\ket{\Psi'} = U^{(j,k)} \ket{\Psi}$
can also be expressed as a UBM-NNS with new vectors:
\begin{equation}
\begin{split}
  \tilde{a} &\!=\! (a_1,\cdots,a_{j-1},\alpha_1,a_{j+1},\cdots,a_{k-1},\alpha_2,a_{k+1},\cdots,a_N)^t\\
  \tilde{b} &\!=\! (b_1,\cdots,b_M,\beta_1 + a_j, \beta_2 + a_k)^t\\
  \tilde{h} &\!=\! (h_1,\cdots,h_M,h_{M\!+\!1},h_{M\!+\!2})^t,
\label{eq:new_ubm_vectors}
\end{split}
\end{equation}
and matrices:
\begin{equation}
\begin{split}
  \tilde{W} &\!\!=\!\!
  \begin{pmatrix}
  |        &          &  0            & |         &          &  0           & |         &          & |       \\
  W_{i,1}  &\!\cdots\!& \vdots        & W_{i,j+1} &\!\cdots\!& \vdots       & W_{i,k+1} &\!\cdots\!& W_{i,N} \\
  |        &          &  0            & |         &          &  0           & |         &          & |       \\
  Y_{j,1}  &\!\cdots\!&  \Omega_{11}  & Y_{j,j+1} &\!\cdots\!&  \Omega_{12} & Y_{j,k+1} &\!\cdots\!& Y_{j,N} \\
  Y_{k,1}  &          &  \Omega_{21}  & Y_{k,j+1} &          &  \Omega_{22} & Y_{k,k+1} &          & Y_{k,N}
  \end{pmatrix},\\
  \tilde{X} &=
  \begin{pmatrix}
            &         &        &         & W_{1,j}      & W_{1,k} \\
            &         &        &         & W_{2,j}      & W_{2,k} \\
            &         &  X     &         & \vdots       & \vdots  \\
            &         &        &         &              &         \\
            &         &        &         & W_{M,j}      & W_{M,k} \\
    W_{1,j} & W_{2,j} & \cdots & W_{M,j} & 0            &\gamma\!+\!Y_{j,k} \\
    W_{1,k} & W_{2,k} & \cdots & W_{M,k} &\gamma\!+\!Y_{j,k}& 0
  \end{pmatrix},\\
  \tilde{Y}_{n,n'} &=
  \begin{cases}
    Y_{n,n'}  \qquad n,n'\neq j,k\\
    \lambda \: \delta_{n',k}  \qquad n=j\\
    \lambda \: \delta_{n',j}  \qquad n=k
  \end{cases}.
\label{eq:new_ubm_matrices}
\end{split}
\end{equation}
Thus, we see that under the action of the two-body operation given by Eq.
{\ref{eq:simple_two_body}}, the resulting NNS can also be described by a UBM
but with two additional hidden nodes.  We have focused on
two-body operations but these results can be directly extended to the case of
$K$-body operations. Thus, we can consider
operations $U$ acting on $K$ two-level subsystems whose
matrix elements $U_{q_1...q_K,q'_1...q'_K} =
\bra{q_1,...q_K} U \ket{q'_1...q'_K}$ can be written as:
\begin{equation}
U_{q,q'} \!=\! A \exp\left(\!
\alpha^t q
\!+\!
\beta^t q'
\!+\! \frac{1}{2} \!
\begin{pmatrix}q^t & q'^t\end{pmatrix} \!\!
\begin{pmatrix}\Lambda & \Omega \\
\Omega^t & \Gamma
\end{pmatrix} \!\!
\begin{pmatrix}q \\ q'\end{pmatrix}\!\!
\right)
\label{eq:simple_K_body}
\end{equation}
where $q=(q_1,\cdots,q_K)^t$, $q'=(q'_1,\cdots,q'_K)^t$, $\alpha$ and
$\beta$ are column vectors with
$K$ components, and $\Lambda$, $\Gamma$ and $\Omega$ are $K\times K$ matrices.
$\Lambda$ and $\Gamma$ are symmetric with null diagonals. Operations in
this family can be applied to any UBM-NNS by adding $K$ hidden nodes and
modifying the offsets and connections weights in a way that is a direct
extension of Eqs. \ref{eq:new_ubm_vectors} and \ref{eq:new_ubm_matrices} for the
case of two-body operations. We refer to operations that can be written as in
Eq. \ref{eq:simple_K_body} as Neural Network Operations (NNOs), since they can
also be represented by a network of nodes or neurons with associated complex
offsets and arbitrary connections between them (with complex weights),
as is depicted in Figure \ref{fig:nqo}. Note that in this case there are no
hidden nodes, although composition of two or more NNOs leads to networks with
hidden nodes.

In Appendix \ref{ap:conditions_unitary} it is shown that an operation $U$ given
by Eq. \ref{eq:simple_K_body} will be unitary if and only if the followings
conditions hold:
(i) The components of $\alpha$,$\beta$,$\Gamma$, and $\Lambda$ are purely
imaginary, (ii) the matrix $\Omega$ should have only one element different from
zero in each row, and (iii) the imaginary part $x$ of each of these elements
should be such that $\cos(2x)=0$. Thus, the number of
independent real parameters is $n_K = K^2+2K$. This should be compared with the
number $m_K = 2^{2K}-1$ of independent real parameters for arbitrary $K$-body
unitaries (apart from global phases). For $K=1$ we have $n_1 = m_1 = 3$, as we
expect since we saw that any one-body unitary is a NNO.
On the other hand, already for $K=2$ we have $n_2=8$
and $m_2=15$. Thus, only a restricted set of two-body unitaries can be cast
as NNO. However, this restricted set includes entangling operations.
As a simple example, we note that for $\alpha = \beta = 0$,  $\Gamma = 0$,
$\Lambda= i\lambda \left(\begin{smallmatrix} 0 & 1\\ 1 & 0\end{smallmatrix}\right)$,
and $\Omega= i(2n+1)\pi/4 \left(\begin{smallmatrix} 1 & 0\\ 0 & 1\end{smallmatrix}\right)$,
we obtain the following two-body unitary:
\begin{equation}
U = \frac{1}{2}
\begin{pmatrix}
  ie^{i\lambda}  &  e^{i\lambda}  &  e^{i\lambda}  & -ie^{i\lambda} \\
  e^{-i\lambda}  &  ie^{-i\lambda}  & -ie^{-i\lambda}  &  e^{-i\lambda} \\
  e^{-i\lambda}  & -ie^{-i\lambda}  &  ie^{-i\lambda}  &  e^{-i\lambda} \\
 -ie^{i\lambda}  &  e^{i\lambda}  &  e^{i\lambda}  &  ie^{i\lambda}
\end{pmatrix},
\end{equation}
which takes product states into maximally entangled states for $\lambda = \pi/4$.
This operation and all the one-qubit operations form a universal set of
quantum gates that can be expressed as NNOs. From this, it follows
that the resulting state of any quantum circuit with $G$ one-qubit and two-qubit
gates can be cast as a UBM-NNS with a number of hidden nodes that
is linear in $G$, provided that the initial state is also a UBM-NNS.
However, we note that the direct application of a $K$-body operation could be
more efficient, in terms of the number of hidden nodes added to the network,
than decomposing it in terms of a set of one-body and two-body primitives.
Thus, it is interesting to investigate what kind of $K$-body operations can
be expressed as NNOs.

As explained in section \ref{sec:one-body}, the action of one-body unitaries that
are diagonal in the computational basis (rotations around $\hat z$) can be
implemented without adding new hidden nodes. The same happens for diagonal
two-body unitaries. Indeed, the controlled rotation
$\exp({-i(\theta/2) \sigma_z^{(j)}\sigma_z^{(k)}})$ can be obtained as a NNO in the
limit $\text{Re}(\Omega) \to +\infty$, and can be implemented without adding hidden nodes
and just updating the matrix $Y$ according to $Y_{j,k} \to Y_{j,k}-i\theta/2$.

\section{Sampling of UBM-NNS\lowercase{s}}
\label{sec:sampling}

The previous results are interesting and promising, but are not directly
useful. In fact, it is not clear how to extract information out
of the representation given by Unrestricted Boltzmann Machines.
In contrast to RBMs, it is not possible to
analytically perform the sum over the hidden nodes of a UBM, since these nodes
interact with each other. Therefore, for general UBM-NNSs, $\Psi(s)$ cannot
be evaluated in an efficient and exact way. Approximate solutions are in
principle possible, but, as we will see, they suffer from the well known
`sign problem' of standard quantum Monte Carlo (QMC)
techniques\cite{troyer2005, loh1990}.

\begin{figure}[ht]
\mbox{
\Qcircuit @C=.7em @R=.7em {
& \gate{g_1} & \multigate{1}{g_2} & \qw & \qw & \qw
& \gate{g_1} & \multigate{1}{g_2} & \qw & \qw & \qw \\
& \gate{g_1} & \ghost{g_2} & \multigate{1}{g_2} & \qw & \qw
& \gate{g_1} & \ghost{g_2} & \multigate{1}{g_2} & \qw & \qw \\
& \gate{g_1} & \qw & \ghost{g_2 }& \multigate{1}{g_2} & \qw
& \gate{g_1} & \qw & \ghost{g_2 }& \multigate{1}{g_2} & \qw \\
& \gate{g_1} & \qw & \qw & \ghost{g_2} & \qw
& \gate{g_1} & \qw & \qw & \ghost{g_2} & \qw \\
}
}
\caption{Approximation of $V(t)$ as a circuit for $N=4$ spins and $S=2$ Trotter steps. }
\label{fig:circuit_ising}
\end{figure}
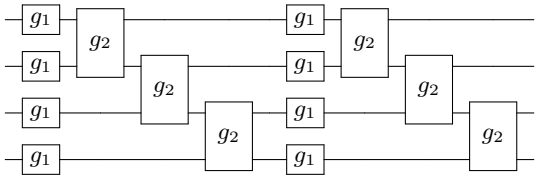

In Eq. (\ref{eq:ubm_nns}) the task is to evaluate the sum over the
variables $h_1,\cdots, h_M$. Without restricting in some way the matrix $X$,
this is at least as hard as computing the partition function
of an arbitrary classical Ising system (which is intractable\cite{barahona1982}).
An approximate numerical solution might be to employ
a Metropolis-like sampling strategy, and only consider the terms of the sum
with larger contributions. However, due to the `sign problem',
this approach can only be applied in a restricted family of problems.
To explain this we will focus on a particular example: the determination
of the ground state of the one dimensional transverse field Ising
model (TFI-1D) via a Susuki-Trotter evolution in imaginary time.
The elementary interactions of this model give rise to operations that
can be easily represented as one-body and two-body NNOs. In this way, we will
be able to explicitly construct a UBM-NNS that approximates the ground state
of the model. The Hamiltonian of this model is:
\begin{equation}
    H = -J\left(\sum_{j=1}^{N-1} \sigma_j^z\sigma_{j+1}^z + h \sum_{j=1}^N
    \sigma_j^x\right)
    \label{eq:tfi_1d}
\end{equation}
where $J>0$, and we consider open boundary conditions.
We want to prepare the ground state of this Hamiltonian via imaginary
time evolution of the following initial state: $\ket{\Psi(0)} =
\otimes_{k=1}^N [(\ket{-1} + \ket{1})/\sqrt{2}]$, which can be considered
as a RBM-NNS with $N$ visible nodes, $M=0$ hidden nodes, $a=0$ and $Y=0$.
This state is `evolved' in imaginary time $t$ with the operator $V(t)=e^{-tH}$.
If $\ket{\Psi(0)}$ has a non-vanishing projection in the ground state subspace,
then $\ket{\Psi(t)} = V(t) \ket{\Psi(0)}$ will belong to that
subspace for $t \to +\infty$. We can approximate the operator $V(t)$ as
a periodic circuit using the first order Susuki-Trotter approximation:
\begin{equation}
V(t) = e^{-tH} = \left(e^{-tH/S}\right)^S \simeq
\left(\prod_{k=1}^{N-1} g_2(k) \prod_{k=1}^{N} g_1(k)\right)^S
\label{eq:trotter_decomp}
\end{equation}
where $S$ is the total amount of steps and
\begin{equation}
\begin{split}
g_1(k)&=e^{\tau Jh \sigma_k^x}\\
g_2(k)&=e^{\tau J \sigma_k^z\sigma_{k+1}^z}
\end{split}
\end{equation}
and  are one-body and two-body elementary operations, respectively
($\tau = t/S$).  A circuit representing this decomposition of $V(t)$ is shown
in Figure \ref{fig:circuit_ising} for $N=4$ and $S=2$.

\begin{figure}[ht]
  \includegraphics{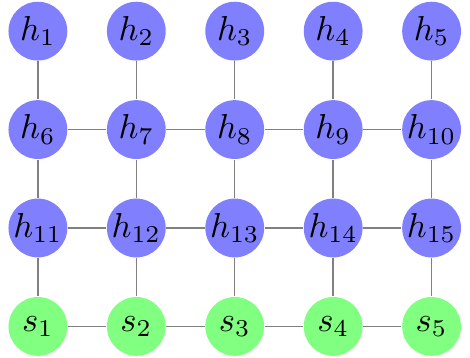}
  \caption{UBM representing $\ket{\Psi(t)} = V(t)\ket{\Psi(0)}$ for $N=5$ spins
   and $S=3$ Trotter steps. The application of each Trotter step adds a layer
   of hidden nodes to the network.}
  \label{fig:ubm_ising}
\end{figure}

Figure \ref{fig:ubm_ising} illustrates the UBM representing the final state
after the application of the decomposition of Eq. (\ref{eq:trotter_decomp})
to $\ket{\Psi(0)}$ for $N=5$ and $S=3$. Since $g_2$ is diagonal in the
computational basis, it can be
implemented without adding hidden nodes, as explained in the previous section.
However, as shown in section \ref{sec:one-body}, each application of $g_1$ adds
one hidden node to the network.
The new hidden nodes organize themselves in a two-dimensional structure with
interactions between first neighbors. The weights of the vertical and horizontal
edges are $w_v = \log(\coth(\tau J h))/2$ and $w_h = \tau J$, respectively.
Of course, we are just recovering the well know correspondence, or duality,
between the TFI-1D model and the 2D classical anisotropic Ising model
\cite{sachdev2007}. We see that the hidden nodes of the UBM
representation of $\Psi(t)$ act as the classical spins of the corresponding
classical model. This can be generalized to more complex models in higher
dimensions.

Now, in order to evaluate the probability amplitude $\Psi(s,t)$ according to
Eq. (\ref{eq:ubm_nns}), we could implement an importance sampling strategy to
numerically approximate the sum. As in usual QMC methods, that
are also based on a quantum-classical correspondence, this will only work
reliably, in principle, if the parameters $b$, $W$ and $X$
in Eq. (\ref{eq:ubm_nns}) are real, such that the factor
$\exp(b^t h+ h^t W s + h^t X h/2)$ is always real and positive for all $h$ and $s$
\footnote{This is in fact the case for the given example, but the imaginary time
evolution in more complex models, or even the real time evolution in the
TFI-1D model, leads to UBMs with complex $X$.}.
Otherwise the numerical `sign problem' will hamper the accurate
estimation of $\Psi(s,t)$\cite{troyer2005, loh1990}.

Thus, so far the situation is completely analogous to the one faced by standard
QMC techniques: a dual classical system is constructed from a quantum
Hamiltonian, and properties of the quantum model are obtained by stochastic
sampling of the classical model, whenever it is free from the sign problem.
However, the representation of quantum states via RBM suggest a way around
this, as we explain in the next section.

\section{Projecting UBM-NNS\lowercase{s} onto RBM-NNS\lowercase{s}}
\label{sec:projection}

As explained in section \ref{sec:one-body},
the sum over hidden nodes of a RBM can be performed analytically. Therefore,
the amplitudes $\Psi(s)$ corresponding to any RBM-NNS can be efficiently
and exactly computed even for complex $b$ and $W$ (see Eq. (\ref{eq:rbm_nns})).
RBM-NNSs are then free from the sign problem, since no importance sampling
is necessary to evaluate $\Psi(s)$. This is true also if the RBM-NNS is allowed to have
interactions between its visible nodes (i.e., if $Y\neq 0$). Thus, in this
section we consider an extended definition of RBM-NNS that allows
for those interactions: a RBM-NNS is just a UBM-NNS with $X=0$.
It should be noted also that the interactions in the visible layer can be
alternatively represented as mediated by additional hidden nodes
(at most $N(N-1)/2$, one for each possible interaction in the visible layer)
\cite{gao2017,glasser2018}.

When a RBM-NNS is subjected to a non trivial evolution,
interactions between hidden nodes will appear and the resulting state
will be described by a UBM-NNS for which, in general, no efficient and accurate
way of computing $\Psi(s)$ is available. One possible approach to avoid this
problem is to continuously project the quantum state back to the family of
RBM-NNSs during its evolution. In this section we explore a possible way to
perform this projection.

\begin{figure}
  \includegraphics{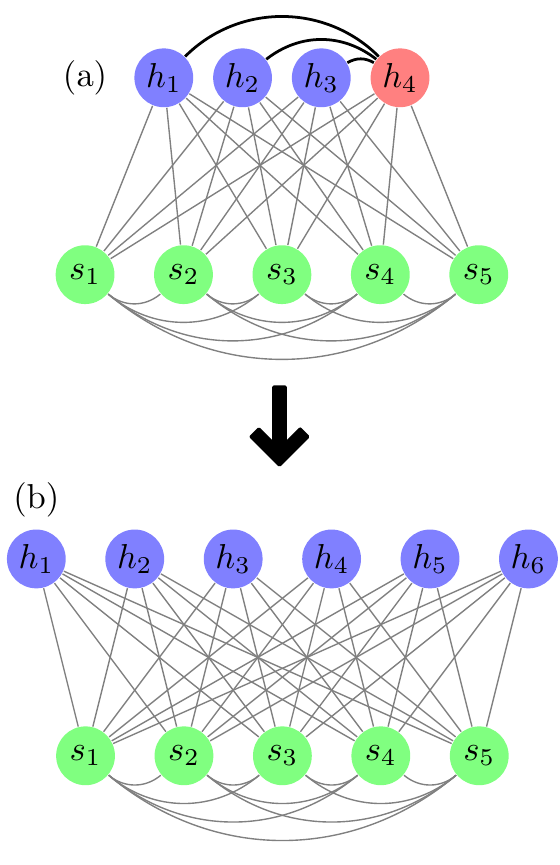}
  \caption{Reduction of UBM-NNS (a) to a RBM-NNS (b). Note that the number of
  hidden nodes in (b) can be in principle larger that in (a), and the fidelity
  of the projection is expected to improve with more hidden nodes.}
  \label{fig:projection}
\end{figure}

In first place, we choose the set of all the one-body unitaries plus the
controlled rotations $\exp({-i(\theta/2) \sigma_z^{(j)}\sigma_z^{(k)}})$ as a
universal set of gates in terms of which we will decompose any global unitary
operation. As we mentioned before, the controlled rotations can be implemented
without adding hidden nodes and \emph{without inducing interactions between the
preexisting hidden nodes}. Therefore, when applying a given evolution
(decomposed as a quantum circuit) to a NNS, hidden nodes will be added to the
network only for one-body operations, and only then will interactions
between the new and preexisting hidden nodes be induced (see Figures
\ref{fig:rbm} and \ref{fig:one_body_ubm}). However, if a one-body NNO is
applied to a RBM-NNS (that, by definition, has no interactions between hidden
nodes), the resulting state will be a UBM-NNS with a very special interaction
structure among its hidden nodes: in principle all hidden nodes will interact
with the newly added hidden node, but not between them. We consider the problem
of projecting this new state back to the RBM-NNS family, as depicted in Figure
\ref{fig:projection}.
Given a procedure to perform this projection,
then by applying it every time an one-body NNO is applied during the
execution of a quantum circuit, the final state after the full evolution will
be an easy to sample RBM-NNS. Of course, any projection procedure is expected to
induce errors, i.e, the fidelity between the original and projected states will
be less that unity. This errors will accumulate during the execution of a
circuit, and this will severely limit the accuracy of the results.

In fact, it is important to note that a general and efficient solution to the
proposed problem is not expected to exist.
Previous works\cite{gao2017} have used complexity theoretical arguments
showing that for many important classes of quantum states, e.g. those
which are efficiently generated by quantum circuits, those which are
representable by PEPS, and those which are ground states of k-local
Hamiltonians, there exist instances which cannot be efficiently represented by RBMs
(exactly nor to to high precision). The existence of a RBM representation of such states would cause
the collapse of the polynomial hierarchy ($PH$) to the third level.
In our approach we, seemingly, attempt to do more: we aim to efficiently
find these representations, given
quantum circuits as input.
The possibility of generically solving such a task has even more
dramatic complexity-theoretic consequences, \emph{e.g.}, if an algorithm
converting between a given circuit, and the RBM representing the output
state to exponential precision which is runnable in polynomial time were
to exist, then $\# P$ problems could be solved in polynomial time as well.
This would imply a
complete collapse of the polynomial hierarchy, \emph{i.e.} $P \supseteq
PH$\footnote{
This holds as $P$ is self-low i.e. $P^P\subseteq P$.
If $\# P$ was solvable in poly-time, then $P \supseteq PP$, and since
$P^{PP} \supseteq PH$ (Toda's Theorem), we have that $P$ contains $PH$.}
, and, in particular $P = NP = BQP$.
However, these arguments do not imply that
no useful states have efficient RBM representations, or algorithms which
construct them. Consequently, any heuristic method which
attempts this, may be to a larger or smaller extent applicable to a given
setting, which is one of the motivations of this work.

Since the structure of the hidden interactions in the states to be reduced is
very simple (see Fig. \ref{fig:projection}-(a)), the sum over hidden nodes in
Eq. \ref{eq:ubm_nns} can still be performed analytically. Indeed, if there are
$M$ hidden nodes and it is considered, without loss of generality, that
the last of them is the one that interacts with all the others, then:
\begin{equation}
\begin{split}
  \Psi(s) =\: & e^{a^t s + s^t Y s/2} \: \times \\
  &\left[ e^{b_M + w_M s} \prod_{k=1}^{M-1} 2 \cosh(w_k s+b_k+X_{k,M}) + \right.\\
  &\left. e^{-b_M - w_M s} \prod_{k=1}^{M-1} 2 \cosh(w_k s+b_k-X_{k,M}) \right],
\end{split}
\label{eq:simple_UBM}
\end{equation}
where $w_k$ is the $k$-th row of $W$, and it was used that only the elements
$X_{k,M}=X_{M,k}$ of $X$ are different from zero. Our goal is to approximate
$\Psi(s)$ with a RBM-NNS $\Psi'(s)$, for which the sum over the hidden nodes
evaluates to:
\begin{equation}
  \Psi'(s) =\: e^{{a'}^t s + s^t Y' s/2} \prod_{k=1}^{M'} 2 \cosh((W's+b')_k),
  \label{eq:red_RBM}
\end{equation}
for a number $M'$ of hidden nodes and new parameters $a'$, $b'$, $W'$, and $Y'$.
In principle it could be $M'\geq M$, since the quality of the approximation is
expected to improve with larger $M'$. In fact, from the previous considerations
about computational complexity, the number $M'$
of new hidden nodes is expected to increase exponentially with $N$ and $M$ in
the general case (for an exact mapping).
Here, we propose two simple methods to perform the approximation of $\Psi(s)$ by $\Psi'(s)$,
and provide numerical evidence in favor of the feasibility of these approaches.
We begin by rearranging the expression in Eq. \ref{eq:simple_UBM} in the
following way:
\begin{equation}
\begin{split}
  &\Psi(s) =\: e^{a^t s + s^t Y s/2} \:\: 2\cosh(\log(\chi(s)))
  \prod_{k=1}^{M-1} 2 f_k(s)
\end{split}
\label{eq:rewrite_simple_UBM}
\end{equation}
where $\chi(s)$ is given by:
\begin{equation}
    \chi(s) = e^{b_M + w_M s} \prod_{k=1}^{M-1}
    \sqrt{\frac{\cosh(w_k s+b_k+X_{k,M})}{\cosh(w_k s+b_k-X_{k,M})}}
\label{eq:def_chi}
\end{equation}
and
\begin{equation}
f_k(s) = \sqrt{\cosh(w_k s\!+\!b_k \!+\!X_{k,M} )\cosh(w_k s\!+\!b_k\!-\!X_{k,M})}
\label{eq:def_fk}
\end{equation}
Equation \ref{eq:rewrite_simple_UBM} has a factored structure similar to a
RBM-NNS wavefunction, as in Eq. \ref{eq:red_RBM}. So far, no approximation has
been made. The first method, explained in the following, consists in giving
appropriate approximations for the functions $\log(\chi(s))$ and $f_k(s)$.

\subsection{First method}

Our goal is to take Eq. (\ref{eq:rewrite_simple_UBM}) to a form comparable to
Eq. (\ref{eq:red_RBM}). A simple way to do that is to propose a linear
approximation for the function $\log(\chi(s))$ and to approximate each factor
$f_k(s)$ as the one corresponding to a single hidden node of a RBM-NNS.
Explictly, we propose to find new vectors $w'_k$ and new constants $b'_k$
such that:
\begin{equation}
f_k(s) \simeq c_k \cosh(w_k's + b'_k)
\label{eq:approx1}
\end{equation}
for each $k=1,\cdots,M-1$, where $c_k$ is an unimportant proportionality factor.
Also, we appoximate $\log(\chi(s))$ as:
\begin{equation}
\log(\chi(s)) \simeq w'_M s + b'_M
\label{eq:approx2}
\end{equation}
for some vector $w'_M$ and offset $b'_M$. In this way, the original state $\Psi(s)$
given by Eq. (\ref{eq:simple_UBM}) is approximated by a RBM-NNS with parameters
$a'=a$, $Y'=Y$, $b'=(b_1',\cdots,b_M')^t$, and a matrix $W'$ with rows
$w'_1,\cdots,w'_M$. In Appendix \ref{ap:first_method} we give details about
the numerical implementation of this method, which is based on the requirement
that the proposed approximations in Eqs. \ref{eq:approx1} and \ref{eq:approx2}
hold exactly when $w_k s = \pm \lVert w_k \rVert_1$ and $w_k s = 0$
($\lVert \cdot \rVert_1$ is the $\ell_1$ vector norm). Here we discuss two
limiting cases in which the proposed approximations can be explicitly found
and are exact.

In first place we consider the limit of strong hidden connections in which
$|X_{k,M}|  \gg | w_k s + b_k | $ for $k=1,\cdots,M-1$.
To first order in this limit (that is, to first
order in $|X_{k,M}|^{-1}$) equation (\ref{eq:approx2}) holds exactly with
parameters $w'_M$ and $b'_M$ given by:
\begin{equation}
  w'_M = w_M + \sum_{k=1}^{M-1} \tanh(X_{k,M}) \: w_k,
\end{equation}
and
\begin{equation}
  b'_M = b_M + \sum_{k=1}^{M-1} \tanh(X_{k,M}) \: b_k,
\end{equation}
while the parameters $b'_k$ and $w'_k$ for $k=1,\cdots,M-1$ vanish. Thus,
in this limit all the hidden nodes in the original state are
condensed into a single one. This is natural, since for large $|X_{k,M}|$ hidden
nodes $M$ and $k$ are highly correlated.

Now we analyze the opposite limit of weak hidden connections, i.e.,
$|X_{k,M}| \ll | w_k s + b_k |$ for $k=1,\cdots,M-1$. This time, to first order in
$|X_{k,M}|$ we have $f_k(s) = \cosh(w_k s + b_k)$. Therefore,
Eq. (\ref{eq:approx1}) holds with $b'_k = b_k$ and $w'_k = w_k$ for
$k=1,\cdots,M-1$. Thus, the first $M-1$ hidden nodes retain their original
parameters. However, also to first order in $|X_{k,M}|$, the function
$\log(\chi(s))$ is given by:
\begin{equation}
\log(\chi(s)) \simeq w'_Ms+b'_M+\sum_{k=1}^{M-1}X_{k,M}\:\tanh(w_ks+b_k)
\end{equation}
Thus, we see that the condition of weak hidden connections is not enough to
assure that a linear approximation for the function $\log(\chi(s))$ holds.
In principle, it is also necessary to assume that the components of $w_k$
are small, so that a linear approximation to each function $\tanh(w_k s+b_k)$
in the sum of the last equation can be given. In particular, to first order
in $w_k$ and $b_k$, we obtain the following expressions for the parameters
of the hidden node $M$.
\begin{equation}
  w'_M = w_M + \sum_{k=1}^{M-1} X_{k,M} \: w_k,
\end{equation}
and
\begin{equation}
  b'_M = b_M + \sum_{k=1}^{M-1} X_{k,M} \: b_k.
\end{equation}
Then, under the conditions mentioned above, the parameters of the hidden node
$M$ in the obtained RBM-NNS are updated by small contributions of the other
hidden nodes connected to it in the original UBM-NNS.

\begin{figure}
  \centering
  \includegraphics[scale=.55]{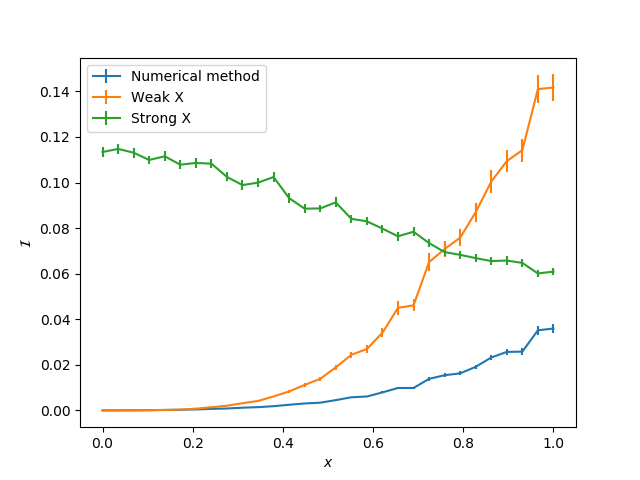}
  \caption{Mean value of the infidelity $\mathcal{I}$ as a function of the $x$,
  for $w=1/5$, $N=12$ and 200 randomly generated states for each point.
  The error bars indicate the uncertainty in the mean values. The blue line
  corresponds to the numerical implementation of the method, while the orange
  and green lines correspond to the projection rules obtained in the limit
  of weak and strong hidden connections, respectively.}
  \label{fig:fidelity}
\end{figure}

We test this projection method on randomly generated NNSs. For this, we consider
NNSs with the same number $N=M=12$ of visible and hidden nodes. The components
of the matrix $W$ (or equivalently, of the vectors $\omega_k$) are selected
from an uniform distribution in the interval $[-w,w]$. The hidden
connections $X_{k,M}$ are selected from an uniform distribution in the
interval $[-x,x]$. All the other parameters ($a$,$b$, and $Y$) are zero.
We fix $w=1/5$ and let $x$ take values between $0$ and $1$.
For each value of $x$, we generate $200$ random NNSs and calculate the average
fidelity $\mathcal{F}$ between each generated state and the one obtained after
the projection. This is done exactly so we limit to $N=12$. Figure
\ref{fig:fidelity} shows the average infidelity $\mathcal{I}=1-\mathcal{F}$
for three different versions of the method: i) the numerical
implementation of the method explained in Appendix \ref{ap:first_method}
ii) the limit of weak hidden connections, and iii) the limit of strong
hidden connections. We see that in the first two cases the infidelity of the
projection is indeed low for small $x$, but the numerical version of the method
is more robust for higher $x$. Also, the infidelity for the projection rules
obtained for strong hidden connections decreases for increasing $x$, as expected.

As a final remark, we note that, as explained at the end of Section
\ref{sec:one-body}, the action of infinitesimal one-body operations
will add hidden nodes to the network, and that the weight of their connections
to the visible layer will increase as the given operations approach the
identity. This compromises the assumption that the components of the vectors
$w_k$ are small, that led to the last couple
of equations. The second method that is presented below attempts to directly
take into account the action of infinitesimal operations.

\subsection{Second method: infinitesimal operations}
\label{sec:method_infinit}

The previous method provides an approximated RBM representation of the UBM-NNS
obtained by the application of a one-body operation to an initial RBM-NNS.
We now present a different method, in which the action of the given operation is
taken into account by just updating the parameters of the initial RBM-NNS.
Consequently, the number of hidden nodes remains constant.
This method only considers the action of infinitesimal one-body operations.
If we apply such an operation to a given RBM-NNS, one might expect to be able
to approximate the resulting state (with reasonable fidelity) by a new RBM-NNS
with slightly different parameters. We will see that this is indeed the case,
at least under some conditions that we discuss in the following.

Thus, we consider a RBM-NNS with parameters $a,Y,b$ and $W$, and the
corresponding wavefunction $\Psi(s)$. If the parameters are modified
as $x'= x + \delta x$ (where $x=a,Y,b\text{ or }W$), then, to first order in the
variations $\delta x$, the new wavefunction $\Psi'(s)$ is given by the
following expression:
\begin{equation}
\begin{split}
\frac{\Psi'(s)}{\Psi(s)} &\simeq 1 + \delta a^t s + s^t \delta Y s/2
+ \delta b^t T(s) + T(s)^t \delta W s \\
&\equiv 1 + \mathcal{P}(s),
\end{split}
\label{eq:psi_delta}
\end{equation}
where the column vector $T(s)$ has components
\begin{equation}
T_j(s) = \tanh(b_j + (Ws)_j).
\label{eq:def_T}
\end{equation}

Now, we consider also a third state, that results from applying an infinitesimal
one-body operation to the original state $\Psi(s)$. We assume for simplicity
that the operation in question is the infinitesimal rotation
$U = e^{-i\delta\theta\sigma_x/2} \simeq  \mathds{1} + A \sigma_x$, with
$A=-i\delta\theta/2$. If this operation is applied to spin $k$, the resulting
wavefunction $\Psi''(s) = \sum_{s'} \bra{s_k} U \ket{s'_k} \Psi(s')$ can be
expressed as:
\begin{equation}
\frac{\Psi''(s)}{\Psi(s)} = 1 + A e^{-2a_k s_k -2 s_k (Ys)_k} C_k D_k(s) \equiv 1+\mathcal{Q}(s),
\label{eq:psi_infinit}
\end{equation}
where the factor $C_k = \prod_{j=1}^M \cosh(2W_{j,k})$ is independent of $s$
and the factor $F_k(s)$ is given by:
\begin{equation}
  D_k(s) = \prod_{j=1}^M \left( 1-T_j(s) \tanh(2W_{j,k}) s_k \right).
\end{equation}

We now compute the fidelity between states $\Psi'(s)$ and $\Psi''(s)$,
given by
$\mathcal{F} = |\braket{\Psi'|\Psi''}|/\sqrt{\braket{\Psi'|\Psi'}\braket{\Psi''|\Psi''}}$.
A rather long but straightforward calculation shows that to first non-trivial
order in $\mathcal{P}$ and $\mathcal{Q}$
(defined in Eqs. (\ref{eq:psi_delta}) and (\ref{eq:psi_infinit})),
the fidelity satisfies:
\begin{equation}
\mathcal{F}^2  = 1 - \text{Var}(\mathcal{P} \!-\! \mathcal{Q}),
\label{eq:fidelity}
\end{equation}
where
$\text{Var}(\mathcal{X}) = \langle (\mathcal{X}^*-\langle \mathcal{X}^*\rangle)
(\mathcal{X} - \langle \mathcal{X}\rangle)\rangle$ and the mean values are
calculated according to the probability distribution given by the original
wavefunction $\Psi(s)$:
\begin{equation}
    \langle \mathcal{X} \rangle = \frac{1}{\sum_{s'} |\Psi(s')|^2}
    \sum_{s} |\Psi(s)|^2 \mathcal{X}(s).
\end{equation}
Thus, $\text{Var}(\mathcal{P} - \mathcal{Q})$ can be interpreted as the expected
variance in the state $\Psi(s)$ of an (in general non-hermitian) operator which
is diagonal in the computational basis, with diagonal
elements $\mathcal{P}(s) - \mathcal{Q}(s)$.

In order to take into account the action of the operation $U$ on spin $k$
by updating the parameters $a,Y,b$ and $W$ of the original state, we should
select the updates $\delta a$, $\delta Y$, $\delta b$ and $\delta W$
that minimize $\text{Var}(\mathcal{P} - \mathcal{Q})$ (and thus maximize the
fidelity $\mathcal{F}$). This is again a complex and highly non-linear optimization
problem. However, as we explain below, it can be easily solved to first order in the parameters
$a,Y$ and $W$.  In that regime, it is possible to select the updates $\delta x$
in such a way that $\mathcal{P}(s) - \mathcal{Q}(s)$ is a constant independent of $s$,
and therefore has no variance (up to the considered order).
This is done as follows.
To first order in $a$ and $Y$ we can approximate the first non trivial factor
appearing in Eq. (\ref{eq:psi_infinit}) as:
\begin{equation}
e^{-2a_k s_k -2 s_k (Ys)_k} \simeq 1 -2a_k s_k -2 s_k (Ys)_k + \cdots
\label{eq:approx_exp}
\end{equation}
Also, to first order in the parameters $W_{j,k}$:
\begin{equation}
D_k(s) \simeq 1 - \sum_{j=1}^M T_j(s) \tanh(2W_{j,k}) s_k  + \cdots
\label{eq:approx_D}
\end{equation}
Introducing these expansions back in Eq. (\ref{eq:psi_infinit}) and comparing the
result to Eq. (\ref{eq:psi_delta}) we can see how to select the updates
$\delta a$, $\delta Y$, $\delta b$ and $\delta W$ in order for
$\mathcal{P}(s) - \mathcal{Q}(s)$ to be independent of $s$. The results are:
\begin{equation}
\begin{split}
\delta a_j &= -2 A C_k \; \delta_{j,k} \; a_k\\
\delta Y_{i,j} &= -2 A C_k \; (\delta_{i,k} Y_{k,j} + \delta_{j,k} Y_{k,i})\\
\delta b_{j} &= 2A C_k \; \tanh(2 W_{j,k}) \; a_k\\
\delta W_{i,j} &=  2A C_k \; \tanh(2 W_{i,k}) \; (Y_{k,j} - \delta_{k,j}/2)
\end{split}
\label{eq:update_rules}
\end{equation}
We stress that the above update rules are only expected to be useful in the
regime where the parameters $a,Y$ and $W$ are sufficiently small. Also, the
perturbative treatment is not consistent, since not
all second order terms were considered but only the bilinear ones.
Proper analytical consideration of higher order terms in Eqs. (\ref{eq:approx_exp}) and
(\ref{eq:approx_D}) might lead to better update rules valid on a wider regime,
althought in that case one also has to face the optimization problem of minimizing
$\text{Var}(\mathcal{P} - \mathcal{Q})$. This will be explored in future works.
In Appendix \ref{ap:est_error} the infidelity $\mathcal{I}$ is explicitly
evaluated to first non-trivial order in $A$ and $W$, for the specific case in
which $a=0$, $b=0$ and $Y=0$.

\subsection{Remarks and comparisons between the two methods}

The numerical implementation of the proposed approximations in
the first method (method \textbf{I}) heavily rely
on the assumption that the parameters $\omega_k$ and $X_{k,M}$ are real.
The projection procedure can still be carried out for complex parameters, but
it is not expected to offer a good approximation of the original state in that
case. Consequently, this method is restricted to models already free from the
sign problem. Be that as it may, this method can still be of practical relevance,
since the RBM representation is more compact and easy to sample than a positive
definite UBM-NNS.

In contrast, the update rules derived for infinitesimal operations
(method \textbf{II}) are in principle valid for real as well for complex
parameters. In the next section we show that they can be used to evolve
states according to the TFI-1D Hamiltonian in real time, where complex
parameters for the instantaneous RBM-NNS are needed.
They are however severely limited by the fact that they were
derived only to first order in the original parameters. This, in turn, limits
the total time to which states can be accurately evolved.

\section{Optimization of RBM-NNS}
\label{sec:optimization}

\subsection{Imaginary time evolution.}
As a proof of concept, we apply the previously introduced ideas and methods
to a simple problem: the approximation with a RBM-NNS of the ground state of the
TFI-1D model. We will compare our results to those originally obtained in
\cite{carleo2017}, where the same model was employed as a testbed, and its ground
state was approximated with a RBM-NNS optimized via a Variational Monte Carlo
algorithm.

We apply the Trotter evolution in imaginary time introduced in Section
\ref{sec:sampling}, now with periodic boundary
conditions. Thus, we repeatedly apply the following Trotter step:
\begin{equation}
S_\tau = \prod_{k=1}^{N} g_2(k) \prod_{k=1}^{N} g_1(k)
\label{eq:trotter_step}
\end{equation}
to the initial state
$\ket{\Psi(0)} = \otimes_{k=1}^N [(\ket{-1} + \ket{1})/\sqrt{2}]$.
Again, $g_1(k)=e^{\tau Jh \sigma_k^x}$
and $g_2(k)=e^{\tau J \sigma_k^z\sigma_{k+1}^z}$ (this time, $N+1=1$ should
be understood), and $\tau$ is the time step (we take $J=1$ in the following).
We will first consider the projection method (\textbf{I}).
Thus, after the application of each $g_1$ operation, we project the resulting
state back to a RBM-NNS according to the procedure explained in
Appendix \ref{ap:first_method}.

After the application of a single Trotter step,
the network representing the state gains $N$ new hidden nodes.
Thus, old hidden nodes could be deleted such that after each step
the factor $M/N$ does not exceeds some integer constant $\alpha$ fixed before
hand. This is the `hidden-variable density' defined in \cite{carleo2017}.
However, the projection method \textbf{I} is such that for this model only the
newest $N$ hidden nodes remain connected to the visible layer, and therefore
with this method the hidden node density $\alpha$ is effectively always $1$.
With more sophisticated projection methods the value of $\alpha$ could be
selected at will.

\begin{figure}
  \centering
  \includegraphics[scale=.55]{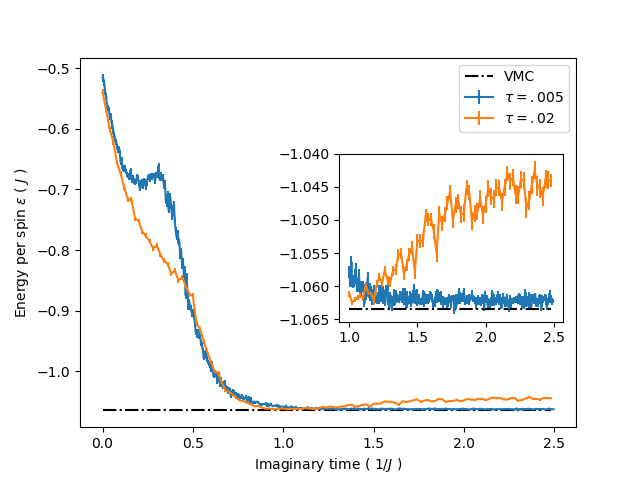}
  \caption{Mean value of the energy per spin as a function of the total imaginary
  time, for two different values of the time step $\tau$. The parameters are
  $N=40$ and $h=0.5$. The dashed black line correspond to the value of energy
  obtained via optimization of an RBM-NNS
  with VMC ($\alpha=1$) \cite{carleo2017_data_40}. The inset shows in detail
  the last part of the optimization. (Projection method \textbf{I}).}
  \label{fig:opt_H_40_.5}
\end{figure}

We first consider an Ising chain with $N=40$ spins and a transverse field $h=0.5$
(which is sub-critical). In Figure \ref{fig:opt_H_40_.5} we show
the mean value of energy per spin $\epsilon = E/N$ for the RBM-NNS obtained
after each Trotter step as a function of the total imaginary time, for different
time steps. The mean value of
the energy $E$ was estimated stochastically in the same
way as in \cite{carleo2017}. We stress that this is only done to monitor the
convergence of the method, but no information from this sampling is used to
assist the optimization, and that the RBM-NNS after any number of
steps can be obtained with no sampling at all. This is in fact the main
advantage of the proposed method.
We observe that in general the energy increases after reaching a minimum value
(this is shown for $\tau =0.02 J^{-1}$ but is also observed for higher values of $\tau$).
However, there are values of $\tau$ for which this `rebound'
is not observed and the energy attains a minimum for large times
(as shown for $\tau \simeq 0.005 J^{-1}$). In this last case, the mean value of
the energy approaches that obtained with VMC \cite{carleo2017_data_40},
but does not improve it.

\begin{figure}
  \centering
  \includegraphics[scale=.55]{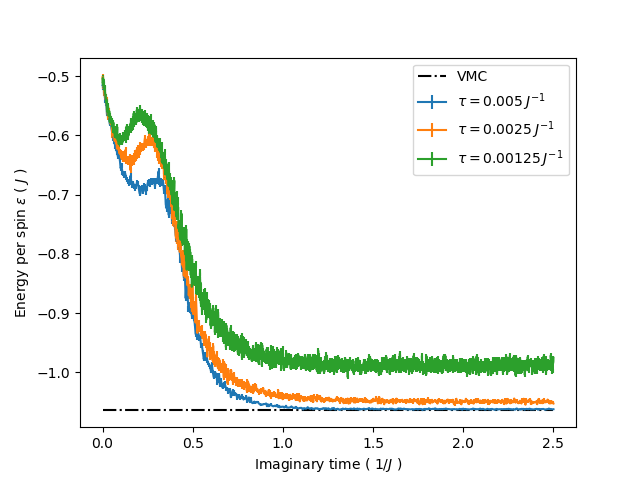}
  \caption{Mean value of the energy per spin as a function of the total imaginary time,
  for decreasing values of the time step $\tau$. The parameters are the same
  that for Figure \ref{fig:opt_H_40_.5}. (Projection method \textbf{I}).}
  \label{fig:opt_H_40_.5_small_step}
\end{figure}

\begin{figure}
  \centering
  \includegraphics[scale=.5]{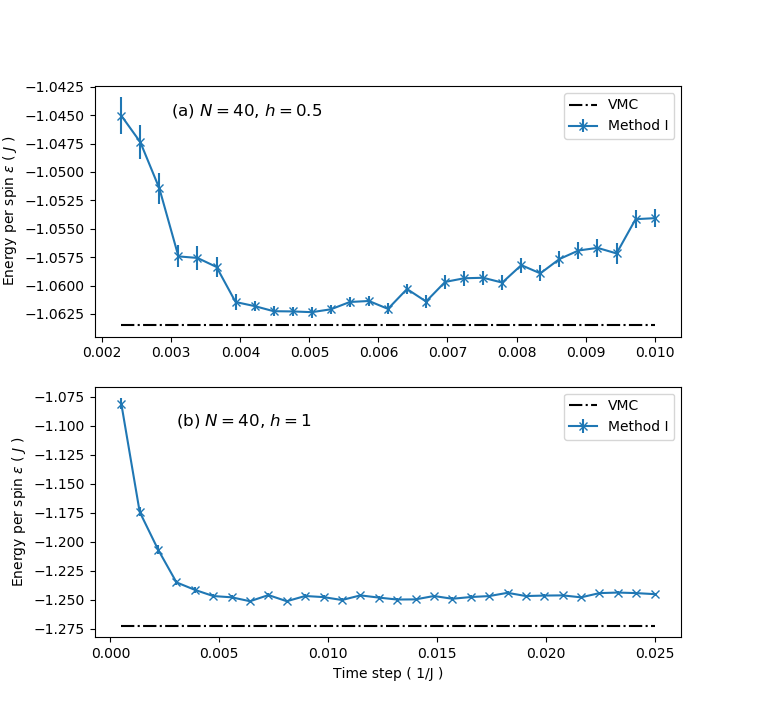}
  \caption{Mean value of the energy per spin as a function of the time step,
  for a total imaginary time  of $t=5$ (a), and $t=6$ (b). The dashed lines
  indicate the energy obtained via optimization of RBM-NNS with
  VMC in each case\cite{carleo2017_data_40,carleo2017_data_40_1}.
  (Projection method \textbf{I}).}
  \label{fig:energy_vs_dt_40}
\end{figure}

Figure \ref{fig:opt_H_40_.5_small_step} shows that not always a smaller
value of the time step $\tau$ produce better results. Thus, there seems to be an
optimal value of $\tau$, for which the energy obtained for large times is
minimum. This is also evident from Figure \ref{fig:energy_vs_dt_40}-(a), where the
asymptotic value of the energy is plotted as a function of $\tau$.
This behavior is due to the fact that
the projection procedure introduces errors that are independent from the
errors introduced by the discretization in the Susuki-Trotter evolution.
It can be qualitatively understood as follows. The errors introduced by the
Susuki-Trotter decomposition increase monotonically with the Trotter step $\tau$.
Therefore, even if the projection method applied after each one-body operation
were free from errors, the fidelity of the generated output states
is expected to decrease as $\tau$ increases. However, for low
values of $\tau$ the error of the projection method dominates
over the error introduced by the Susuki-Trotter discretization. This is so,
as discussed in Section \ref{sec:one-body}, since the action of infinitesimal
one-body operations requires strong connections between the hidden and visible
layers. This, as it follows from the analysis of Section \ref{sec:projection}
for weak hidden connections, is expected to reduce the fidelity of the projection
method. Thus, there is a trade-off between these two sources of errors that
determines an optimal value of the Trotter step $\tau$.

Figure \ref{fig:energy_vs_dt_40}-(b) shows that for the critical case in
which $h=1$ the energy gap between our solution and that obtained with VMC
increases. This fact might point out to some limitation of projection
method \textbf{I} to deal with the long-range correlations present in the critical
ground state, although this state has proven to be harder to approximate even
with variational methods\cite{carleo2017}.

Now we turn to the consideration of the projection method \textbf{II} developed
for infinitesimal one-body operations in Section \ref{sec:method_infinit}.
In this case, we take into account the action of each one-body operation by
just updating the parameters of the RBM-NNS according to
Eqs. \ref{eq:update_rules}, without adding any hidden node. Thus, we need to
provide an initial state with the desired number of hidden nodes.
To motivate the choice of the initial state that is used in the following,
we note that the two-body operations $g_2(k)=e^{\tau J \sigma_k^z\sigma_{k+1}^z}$
can be exactly taken into account
by adding a
hidden node $h$ which is connected to the visible nodes $k$ and $k+1$ with strength
$w = W_{h,k} = W_{h,k+1} = \text{arccosh}(e^{2\tau J})/2$.
Thus, we take as initial state a RBM-NNS with $a=0$, $b=0$, $Y=0$ and
a matrix $W$ with components $W_{h,v} = w (\delta_{h,v} + \delta_{h+1,v})$,
for $h=1,\cdots,N$ and $N+1 \to 1$.
Then, the initial state has $M=N$ hidden nodes ($\alpha=1$), each of which
is connected with the same strength to two successive visible nodes, as if the
operations $g_2(k)$ ($k=1,\cdots,N$) were already applied once to the state
$\ket{\Psi(0)}$ employed previously.

\begin{figure}
  \centering
  \includegraphics[scale=.55]{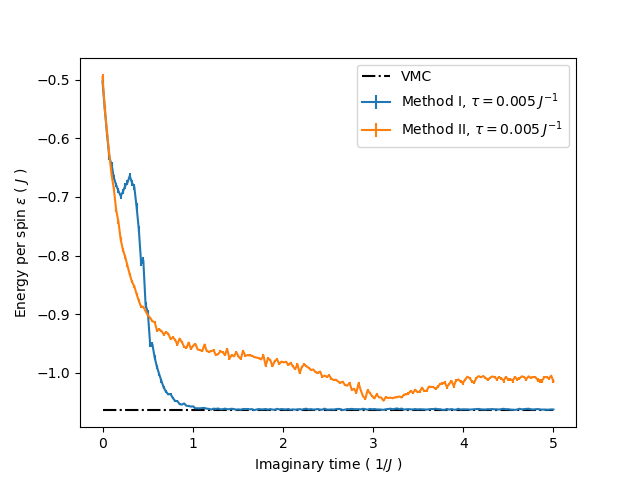}
  \caption{Convergence of the energy optimization for the two projection
  methods \textbf{I} and \textbf{II}. The parameters are the same as in the
  Figure \ref{fig:opt_H_40_.5}.}
  \label{fig:opt_H_vs_I}
\end{figure}

Figure \ref{fig:opt_H_vs_I} compares the convergence of method \textbf{I} and \textbf{II}.
The two methods seems to follow the same curve for short times,
although method \textbf{I} attains lower energies than method \textbf{II}
for later times. In fact, for this last method the energy increases after
reaching a minimum value.

We note that although the proposed methods are not able to improve the
results obtained by Variational Monte Carlo, they can still be employed to
efficiently obtain partially optimized states that can be afterward refined by
stochastic methods. In this way, the total computational effort might be
reduced, in comparison to a fully stochastic optimization.

\subsection{Real time evolution.}
In all the above examples it was only necessary to deal with real parameters
$a,b,Y$ and $W$ during the optimization of the RBM-NNSs. In fact,
to study the ground state of the one-dimensional TFI model, we could have also
sampled a `two-dimensional' positive definite UBM-NNS wavefunction, as
explained in Section \ref{sec:sampling}, since this model is free from the sign
problem. To test if the projection method \textbf{II} is capable of dealing with
RBM-NNSs with complex parameters, we study the real time evolution of the same
model. In this case, complex parameters are necessary to track the time
evolution.

Thus, we consider the Susuki-Trotter decomposition of the unitary operator
$U(t) = e^{-itH}$, which is the same as before (Eq. \ref{eq:trotter_step}),
this time in terms of elementary unitaries $g_1(k) = e^{i\tau Jh \sigma_k^x}$
and $g_2(k) = e^{i\tau J\sigma_k^z\sigma_{k+1}^z}$. As before we take an
initial RBM-NNS with $a=0$, $b=0$, $Y=0$ and a matrix $W$ with components
$W_{h,v} = w (\delta_{h,v} + \delta_{h+1,v})$, for $h=1,\cdots,N$ and $N+1 \to 1$.
This time, however, we must take $w = \text{arccosh}(e^{i2\tau J})/2$.
This initial state is (for small $\tau$) a good approximation of the ground state
of the considered model for $h\to+\infty$. We then evolve it in time with $h=2$ or
$h=1/5$ and measure the expectation value $\langle \sigma^x \rangle$ of the
magnetization in the transverse direction as a function of time.

\begin{figure}[ht]
  \centering
  \includegraphics[scale=.55]{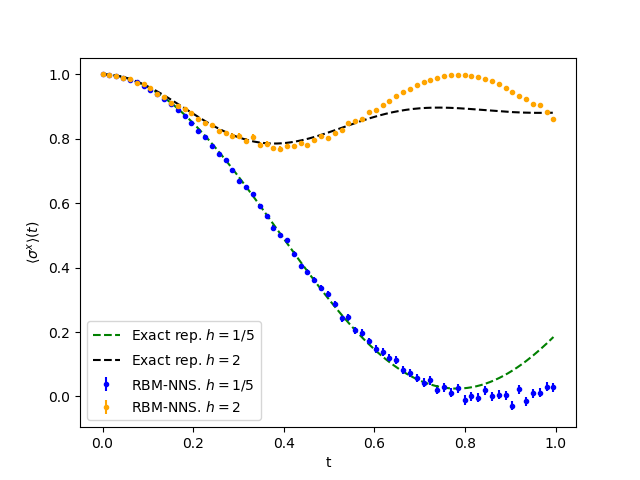}
  \caption{Evolution of $\langle \sigma^x \rangle$ for the two instantaneous
  quenches $h=+\infty \to 2$ and $h=+\infty \to 1/5$. The number of spins is
  $N=24$ and the Trotter step is $\tau =0.005 J^{-1}$. Dashed lines correspond to the
  exact expectation value of $\sigma^x$ obtained from the exact representation
  of the wavefunction, while dots indicate the stochastic estimates for the
  same quantity obtained by sampling the RBM-NNSs.}
  \label{fig:sigmax_vs_t}
\end{figure}

We compared the results with those obtained from the same Susuki-Trotter
evolution but using an exact representation of the wavefunction, to which
the operations $g_1(k)$ and $g_2(k)$ are also applied exactly. For this
reason we limit to $N=24$. Figure \ref{fig:sigmax_vs_t} shows the evolution
of $\langle \sigma^x \rangle$ for the two quenches and the two representations
considered. For short times, the results obtained using RBM-NNSs and the
update rules of method \textbf{II} qualitatively agree with the ones obtained
from the exact representation.

\section{Discussion}

In this work we considered the problem of evolving or optimizing Neural Network
States based on Restricted Boltzmann Machines
without relying on stochastic methods. Our aim was to identify simple and efficient
update rules to take into account the action of elementary operations on
quantum states, on the level of the neural network representation of those states.
We showed that the application of very simple one-body unitaries to
Neural Network States (NNSs) based on Restricted Boltzmann Machines
(RBMs) motivates an extension of this class in order to include NNSs based on
Unrestricted Boltzmann Machines (UBMs). We have parametrized a family of
$K$-body operations that can be efficiently applied to states in this new class.
We showed that there are universal quantum gates included in this
family, and therefore that the action of
any quantum circuit on a NNS can be efficiently represented by a new NNS with
a number of new hidden nodes that grows linearly with the number of elementary
operations in the circuit. This results are similar to the ones
obtained recently in \cite{gao2017}. However,
we give a more general parametrization of many body operations,
which offer more freedom to choose a set of universal quantum gates in
terms of which to decompose general quantum circuits.
We also showed that the action of one-body rotations and two-body controlled
rotations that are diagonal in the computational basis can be taken into
account without leaving the family of RBM-NNSs.

As an application of our study of quantum operations, we investigated a
procedure to optimize or evolve RBM-NNSs in such a way that the evolved state
is still parametrized in this way. This procedure is based on the solution of
a basic problem involving Boltzmann Machines: the reduction to a RBM
of an UBM with only a single hidden node connected to the others.
Two approximate methods to perform
this reduction or projection were discussed. As a proof of concept, we applied
these methods to the imaginary and real time evolution of the transverse
field Ising model in one dimension. In the case of imaginary time evolution we
compared our results with those obtained via Variational Monte Carlo (VMC) methods.
Although the quality of the results offered by our methods is not comparable
to the solutions obtained with VMC, we think that the proposed methods could
be useful as a first stage in a global optimization. However, before applying
these methods to more complex problems, it is necessary to perform a deeper study
of the their properties and limitations, in particular of the errors that they
introduce. More elaborate and accurate projection methods might also be
developed. In this work we limited ourselves to show that deterministic
(as opposed to stochastic) optimization or evolution of RBM-NNSs is in
principle possible.

While investigations of this kind are pragmatically motivated
- we aim to find more efficient methods for computational physics -
they could also establish connections between machine learning,
computational complexity theory, and quantum physics.
Further, they can serve as inspiration for novel quantum algorithms.
For instance, we note that any UBM-NNS which can be reached starting
from some simple initial state using a polynomial circuit is also a UBM whose
distribution can be efficiently sampled on a quantum computer
(but, likely, not classical). Therefore, this offers a route for identifying
quantum algorithms for sampling Unrestricted of Deep Boltzmann Machines.

\section{Acknowledgements}
We would like to thank Ivan Glasser, Nicola Pancotti, Matthias Hein
and Antoine Gautier for useful discussions.

\bibliographystyle{unsrt}
\bibliography{references.bib}

\appendix
\section{Unitary $K$-body NNOs}
\label{ap:conditions_unitary}
In this appendix we give sufficient and necessary conditions for a $K$-body
Neural Network Operation to be unitary. As explained in the main text, these
operations are defined with respect to a given computational basis
$\{\ket{q_1,...,q_K}\}$ as those whose
matrix elements $U_{q_1...q_K,q'_1...q'_K} =
\bra{q_1,...q_K} U \ket{q'_1...q'_K}$ can be written as:
\begin{equation}
U_{q,q'} \!=\! A \exp\left(\!
\alpha^t q
\!+\!
\beta^t q'
\!+\! \frac{1}{2} \!
\begin{pmatrix}q^t & q'^t\end{pmatrix} \!\!
\begin{pmatrix}\Lambda & \Omega \\
\Omega^t & \Gamma
\end{pmatrix} \!\!
\begin{pmatrix}q \\ q'\end{pmatrix}\!\!
\right)
\label{eq:ap_k_body_nno}
\end{equation}
where $q=(q_1,\cdots,q_K)^t$, $q'=(q'_1,\cdots,q'_K)^t$, $\alpha$ and
$\beta$ are column vectors with
$K$ components, and $\Lambda$, $\Gamma$ and $\Omega$ are $K\times K$ matrices.
$\Lambda$ and $\Gamma$ are symmetric with null diagonals.

If the matrix $U$ is unitary then it must be $U^\dagger U = U U^\dagger = \mathds{1}$.
From Eq. \ref{eq:ap_k_body_nno} the diagonal elements of $U^\dagger U$ are:
\begin{equation}
\begin{split}
(U^\dagger U)_{q,q} \!=\! |A|^2 \exp\left(
(\beta+\beta^*)^t q + \frac{1}{2} q^t (\Gamma + \Gamma^*) q \right) \times \\
\sum_r \exp\left( (\alpha + \alpha^*) r  + r^t(\Omega + \Omega^*)q
+ \frac{1}{2} r^t(\Lambda + \Lambda^*) r \right)
\end{split}
\end{equation}
where $r=(r_1,\cdots,r_K)^t$. If $U^\dagger U = \mathds{1}$, then the previous
expression should be independent of $q$. For this to happen it is clear that
it should be $\text{Re}(\beta) = \text{Re}(\Gamma) = 0$. Applying the same condition
to $U U^\dagger$ it follows that $\text{Re}(\alpha) = \text{Re}(\Lambda) = 0$.
Therefore, the previous expression can be reduced to:
\begin{equation}
\begin{split}
(U^\dagger U)_{q,q} &= |A|^2
\sum_r \exp\left(  r^t(\Omega + \Omega^*)q \right) \\
&=\prod_{i=1}^K 2 \cosh\left(\sum_j (\Omega_{i,j} + \Omega_{i,j}^*)q_j\right)
\end{split}
\end{equation}
This last expression will be independent of $q$ if and only if the matrix
$\text{Re}(\Omega)$ has at most one element per row different from zero (since each
component of $q$ is just $\pm 1$ and $\cosh(x)$ is an even function).
Finally, the non-diagonal elements of $U^\dagger U$ are:
\begin{equation}
\begin{split}
& (U^\dagger U)_{q,s} = |A|^2 \exp\left(
\beta^t (s\!-\!q) + \frac{1}{2} (s^t \Lambda s\!-\!q^t \Lambda q) \right) \times \\
& \prod_{i=1}^K\! 2 \cosh\!\!\left(\!\frac{1}{2}\!\sum_j (\Omega_{i,j} \!+\! \Omega_{i,j}^*)
(q_j\!+\!s_j)
+ (\Omega_{i,j} \!-\! \Omega_{i,j}^*)(s_j\!-\!q_j)
\!\!\right).
\end{split}
\end{equation}
Now, since $\text{Re}(\Omega)$ has only one element different from zero in each
row, and since we are assuming $q\neq s$, at least one of the factors in the
last line of the last equation is equal to:
\begin{equation}
 2 \cosh\!\!\left(\!\frac{1}{2}\!\sum_j (\Omega_{i,j} \!-\! \Omega_{i,j}^*)(s_j\!-\!q_j)
\!\!\right).
\end{equation}
This factor will always vanish if $2\times \text{Im}(\Omega)$ has also only one element
different from zero in each row, in the same positions of the non-zero elements
of $\text{Re}(\Omega)$, and each of them is such that it cosine vanishes.

\section{Implementation of the first projection method}
\label{ap:first_method}

In this Appendix we give details about the numerical implementation of
the first projection method.

\emph{First part.}
Given $w_k$, $b_k$, and $X_{k,M}$ for $1\leq \! k \! \leq \! M\!-\!1$, we need
to find $c_k$, $w'_k$ and $b'_k$ such that the factor $f_k(s)$ defined
in Eq. (\ref{eq:def_fk}) is approximated by $c_k\cosh(w'_k s+b'_k)$, as in Eq.
(\ref{eq:approx1}).
We consider $w'_k=\beta w_k$, for some constant $\beta$, so we just need to
determine the constants $c_k$, $\beta$ and $b'_k$. We choose them imposing that
the right and left hand side of Eq. (\ref{eq:approx1}) coincide for
$ws=0$, $ws= \lVert w \rVert_1$, and $ws= -\lVert w \rVert_1$ ($\lVert \cdot \rVert_1$
is the $\ell_1$ vector norm). This is done numerically with an iterative method.

\emph{Second part.}
We need now to give a linear approximation of the function $\log(\chi(s))$.
From Eq. \ref{eq:def_chi}, we have:
\begin{equation}
  \log(\chi(s)) = b_M+w_M s+\frac{1}{2} \sum_{k=1}^{M-1} g_k(s)
\end{equation}
where we define $g_k(s)= \log(\cosh(w_k s + b_k + X_{k,M})) -
\log(\cosh(w_k s + b_k - X_{k,M}))$ for each $1\leq k \leq M\!-\!1$.
We approximate each of these functions as $g_k(s) \simeq \alpha_k w_k s + b_k$, and
the constants $\alpha_k$ and $\beta_k$ are obtained by requiring the approximation
to be exact for $w_k s = \lVert w \rVert_1$ and $w_k s = -\lVert w \rVert_1$.
Thus, the new parameters for the hidden node $M$ are
$b'_M = b_M + \sum_{k=1}^{M-1} \beta_k$ and $w'_M = w_M + \sum_{k=1}^{M-1} \alpha_k w_k$.

\section{Estimation of the error for infinitesimal operations}
\label{ap:est_error}

In this section we evaluate the expression in Eq. (\ref{eq:fidelity}) for the projection
fidelity of method \textbf{I} in a particular case. We consider that the original
state has parameters $a=0$, $b=0$, and $Y=0$, and we will analyze how the
infidelity $\mathcal{I} = 1 - \mathcal{F}$ scales for small $W$. From Eq.
(\ref{eq:fidelity}) it is clear that to first order in $\mathcal{P}$ and
$\mathcal{Q}$ (defined in Section \ref{sec:method_infinit}) the infidelity
satisfies:
\begin{equation}
\mathcal{I} \simeq \text{Var}(\mathcal{P} - \mathcal{Q})/2,
\end{equation}
where
$\text{Var}(\mathcal{X}) = \langle (\mathcal{X}^*-\langle \mathcal{X}^*\rangle)
(\mathcal{X} - \langle \mathcal{X}\rangle)\rangle$ and the mean values are
calculated according to the probability distribution given by the original
wavefunction $\Psi(s)$:
\begin{equation}
    \langle \mathcal{X} \rangle = \frac{1}{\sum_{s'} |\Psi(s')|^2}
    \sum_{s} |\Psi(s)|^2 \mathcal{X}(s)
\end{equation}

For the particular case mentioned above ($a=0$,$b=0$, $Y=0$ and a given $W$),
we obtain the following expression for
$\mathcal{X}(s)=\mathcal{P}(s) - \mathcal{Q}(s)$:
\begin{equation}
\mathcal{X}(s)\! = \! AC_k \! \left[1\!+\!
\sum_{m,n} \! T_m(s) T_n(s) \tanh(2W_{m,k}\!)\tanh(2W_{n,k}\!)\! \right]
\end{equation}
where, as defined in Eq. (\ref{eq:def_T}), $T_n(s)=\tanh((Ws)_n)$ (recall that
we are considering $b=0$). When the components of $W$ are sufficiently small
we can approximate $T_n(s) \simeq (Ws)_n$ and therefore:
\begin{equation}
\mathcal{X}(s) =  AC_k \left[1 + 4
\sum_{i,j} (W^t W)_{i,k}\: (W^t W)_{j,k} \: s_i s_j \right]
\end{equation}
Now we note that for $W\to 0$ the distribution $|\Psi(s)|^2$ is completely flat,
i.e, all configurations $s$ are equiprobable, and as a consequence we have
$\langle s_i s_j \rangle \simeq \delta_{i,j}$. Thus we can readily evaluate
$\langle \mathcal{X} \rangle$ for small $W$, obtaining:
\begin{equation}
\langle \mathcal{X}(s) \rangle =  AC_k \left[1 + 4
(W^t W W^t W)_{k,k}  \right]
\end{equation}
Also, in the same limit we have $\langle s_i s_j s_k s_l\rangle =
\delta_{j,k}\delta_{l,m}+\delta_{j,l}\delta_{k,m}+\delta_{j,m}\delta_{k,l}
-2\delta_{j,k,l,m}$, an we can use this identity to evaluate
$\langle  \mathcal{X}^* \mathcal{X} \rangle$. In this way we arrive at the
final result:
\begin{equation}
\begin{split}
& \mathcal{I} = \text{Var}(\mathcal{X})/2= (\langle \mathcal{X}^* \mathcal{X} \rangle -
\langle \mathcal{X}^* \rangle \langle \mathcal{X} \rangle)/2 \\
&=16 |A C_k|^2 \sum_{i\neq j} (W^t W)_{i,k}(W^t W)_{i,k}^*
(W^t W)_{j,k} (W^t W)_{j,k}^*
\end{split}
\end{equation}
From this expression we see that if $\lambda$ is the typical scale of the
components of the matrix $W$, the infidelity scales as $\mathcal{I} \propto
|A|^2 \lambda^8 N^2$. This scaling with $N$ corresponds actually to the worst
case scenario in which there is no notion of locality in the matrix $W$ (i.e,
a given hidden neuron is in principle connected to all visible nodes, and not
only to a group of them of restricted size).

\end{document}